# Direct Evidence for the Source of Reported Magnetic Behavior in "CoTe"

Zhiwei Zhang,[1,2] Joseph I. Budnick,[1] William A. Hines,[1] David M. Perry,[1] Barrett O. Wells[1,2,*]

[1]Department of Physics, University of Connecticut, Storrs, CT 06269-3046, USA
[2]Institute of Materials Science, University of Connecticut, Storrs, CT 06269-3136, USA

ABSTRACT

In order to unambiguously identify the source of magnetism reported in recent studies of the Co-Te system, two sets of high-quality, epitaxial $CoTe_x$ films (thickness ≈ 300 nm) were prepared by pulse laser deposition (PLD). X-ray diffraction (XRD) shows that all of the films are epitaxial along the [001] direction and have the hexagonal NiAs structure. There is no indication of any second phase metallic Co peaks (either fcc or hcp) in the XRD patterns. The two sets of $CoTe_x$ films were grown on various substrates with PLD targets having Co:Te in the atomic ratio of 50:50 and 35:65. From the measured lattice parameters c = 5.396 Å for the former and c = 5.402 Å for the latter, the compositions $CoTe_{1.71}$ (63.1% Te) and $CoTe_{1.76}$ (63.8% Te), respectively, are assigned to the principal phase. Although XRD shows no trace of metallic Co second phase, the magnetic measurements do show a ferromagnetic contribution for both sets of films with the saturation magnetization values for the $CoTe_{1.71}$ films being approximately four times the values for the $CoTe_{1.76}$ films. $^{59}Co$ spin-echo nuclear magnetic resonance (NMR) clearly shows the existence of metallic Co inclusions in the films. The source of weak ferromagnetism reported in several recent studies is due to the presence of metallic Co, since the stoichiometric composition "CoTe" does not exist.

*wells@phys.uconn.edu

I.    INTRODUCTION

Metal tellurides have been the focus of considerable research activity recently due to their unique properties and vast potential for practical applications. In particular, the Co-Te system, expressed here as $CoTe_x$, is being investigated for use as a non-precious metal electrocatalyst as well as for other specialized materials [1]. $CoTe_x$ has been synthesized with a variety of nanostructured morphologies [1-5] and there are reports in the literature that they are magnetic semiconductors with distinctive electrical transport properties [2,4]. In spite of this effort, there are still differences of opinion concerning the nature of the magnetic behavior for $CoTe_x$. In a very early study, Uchida [6] reported that $CoTe_x$, with $1.00 \leq x \leq 1.20$ was ferrimagnetic with the "stoichiometric CoTe" compound having a saturation magnetization of 7.52 emu/g (0.25 $\mu_B$/Co) and ordering temperature of 1003 °C. Furthermore, when $x = 1.20$, the compound was no longer magnetic and became weakly paramagnetic with no temperature dependence. A short time later, Uchida [7] made detailed magnetic measurements on $CoTe_x$ samples with $0 \leq x \leq 1.00$, and concluded that the magnetic behavior observed earlier could be explained by assuming a eutectic mixture of metallic cobalt and the nonmagnetic compound $CoTe_{1.20}$. In a later Mössbauer study, Fano and Ortalli [8] concluded that the stoichiometric composition CoTe does not exist. However, there are still several recent reports in the literature of Co-Te nanostructured materials having the CoTe stoichiometry and being magnetic [9-13]. One such study reports CoTe nanowires which exhibit ferrimagnetic behavior well above room temperature with a saturation magnetization of 0.2 $\mu_B$/Co [10]. Another report concerning CoTe nanotubes shows hysteresis loops with coercivity indicating ferromagnetic behavior [11]. A magnetic transition from paramagnetism to ferrimagnetism at approximately 40 K has been seen in Co-rich $CoTe_{0.79}$ nanostructures [12]. Finally, a very recent study reports weak ferromagnetism in 100 nm nanorods [13]. As discussed below, an explanation of the magnetic behavior for $CoTe_x$ follows from an understanding of the phase diagram and the nuclear magnetic resonance (NMR) results reported here.

Figure 1 shows the equilibrium phase diagram for the binary Co-Te system based on a review of available data assembled by K. Ishida and T. Nishizawa [14]. The Co-rich region has the ε-Co terminal phase with negligible solid solubility of Te and is stable up to 422 °C. The hexagonal β(Co$_2$Te$_3$) phase has the NiAs structure (space group P6$_3$/mmc) with a homogeneity range from 54.5 at.% Te (CoTe$_{1.20}$) to 64.3 at.% Te (CoTe$_{1.80}$), and is stable up to the congruent melting temperature of 1050 °C. (There is some variation in the reports for the upper limit of this phase, e.g., 65.0 at.% Te [14].) As discussed above, Uchida [7] has suggested that the Co-Te system only becomes phase homogeneous at CoTe$_{1.20}$. Alloys with lower Te content contain the β-phase and metallic cobalt. The orthorhombic phase γ(CoTe$_2$) (space group Pnnm) is homogeneous over the range 66.5 at.% Te (CoTe$_{1.99}$) to 70 at.% Te (CoTe$_{2.33}$) and stable up to the peritectic temperature of 764 °C. One interesting feature of the β(Co$_2$Te$_3$) phase is that as the Te content increases from $x = 1.20$ to 1.80, Co atoms are removed continuously from the middle layer (00½) in the unit cell (see insert in Fig. 1) with the lattice parameters also varying continuously from a = 3.8937 Å, c = 5.3763 Å to a = 3.8017 Å, c = 5.4094 Å, respectively [14,15]. As the Te content increases further, the system passes through a narrow two phase region and evolves into the orthorhombic γ(CoTe$_2$) phase.

In this report, the structural and magnetic properties of two sets of CoTe$_x$ epitaxial 300 nm films grown by pulse laser deposition (PLD) on different substrates are presented. The principal motivation for growing relatively thick films was to create a pathway for exploring the properties for high-quality bulk CoTe$_x$ materials, not "thin" film properties. Epitaxial film growth is a way of obtaining single-crystal-like materials. The films, with Te content $1.20 \leq x \leq 1.80$, consisted of the β(Co$_2$Te$_3$) phase as the principal phase; however, $^{59}$Co spin-echo nuclear magnetic resonance (NMR), along with complimentary magnetic measurements, clearly showed small amounts of a metallic Co second phase. Due to an enhancement factor which occurs for ferromagnetically-ordered materials, NMR can detect the existence of trace amounts of a second magnetic phase which are not seen in x-ray diffraction (XRD). In addition, NMR can distinguish between two phases which have very similar XRD patterns. The results presented here clearly identify metallic Co as the source of magnetism observed in several recent studies. Furthermore, although the Co-Te system has been studied for several decades, there is still a misunderstanding of the phase diagram concerning the existence of the stoichiometric composition CoTe.

## II. EXPERIMENTAL

### A. Sample Preparation

The $CoTe_x$ films were grown using PLD with a target consisting of elemental cobalt and tellurium powders. The powders were mixed together and shaken vigorously. A thickness value of ≈ 300 nm was obtained from the cross section images of a representative film using scanning electron microscopy (SEM). The commercial substrates, which included MgO, $SrTiO_3$ (STO), and $Al_2O_3$, were all 0.5 mm thick. The first set of $CoTe_x$ films, designated CoTe(1), was grown with a Co:Te target mixture having the atomic ratio 50:50, while the second set, designated CoTe(2), was grown with a Co:Te target mixture of 35:65. From previous work by Narangammana et al. [16] on FeTe, another binary telluride compound, a substrate growth temperature of 360 °C was used with a growth pressure of $4.0 \times 10^{-7}$ torr. The substrate-target distance was 4 cm. The 248 nm KrF laser pulses with power 4.8 watt/cm$^2$ impacted the target with a frequency of 4 Hz. The target rotates at 1 Hz during the deposition.

### B. X-Ray Diffraction

X-ray diffraction was carried out on the $CoTe_x$ films using a 2-circle Bruker D2 Phaser diffractometer and a 4-circle Oxford XCalibur$^{TM}$ PX Ultra (2D) diffractometer. Four representative films are described below which highlight the results of this study; they are CoTe(1)-MgO, CoTe(1)-$Al_2O_3$, CoTe(1)-STO, and CoTe(2)-MgO. The XRD patterns obtained from the Bruker diffractometer are shown Fig. 2(a). For all four of the epitaxial films, (001), (002), and (004) peaks were seen and assigned to the NiAs structure which characterizes the $\beta(Co_2Te_3)$ phase [15]. As discussed below, confirmation that these reflections can be assigned to the NiAs structure was obtained from the Oxford 2D diffractometer XRD patterns. The 2θ-values for the peak positions were essentially independent of the substrate; however, there was a small difference between the CoTe(1)-MgO and CoTe(2)-MgO films as shown in Fig. 2(b). The measured lattice parameters were c = 5.396 Å for the former and c = 5.403 Å for the latter. Using the dependence of the c lattice parameter on the Te content x provided by Haraldsen et al. [15], the compositions $CoTe_{1.71}$ and $CoTe_{1.76}$ are assigned to the CoTe(1) and CoTe(2) films, respectively. These compositions are within the range for the $\beta(Co_2Te_3)$ phase, $1.20 \leq x \leq 1.80$. Concerning the XRD patterns shown in Figs. 2(a) and 2(b), two points should be made. (1) There was no indication of any second phase metallic Co peaks in the XRD patterns. (2) With a complete hexagonal NiAs structure, the (00$\ell$) peaks where $\ell$ is odd are forbidden. However, with the removal of some Co in the middle plane (see Fig. 1, insert), these reflections will start to appear. This can explain the relatively weak appearance of the (001) peak in the patterns shown in Fig. 2(a); however, the (003) peak is still not seen. A possible explanation is that it is too weak to be detected in the XRD patterns [17].

Figure 3(a) shows the diffraction image for the CoTe(1)-$Al_2O_3$ film obtained using a 2D detector in a $\theta$-scan with the 4-circle Oxford diffractometer. The image was obtained with the film aligned such that the c-axis stayed in the same plane as the incoming beam during sample rocking. All of the spots are labeled with the corresponding reciprocal lattice G-vector values for the hexagonal NiAs ($P6_3/mmc$) crystal structure [18]. Figure 3(b) shows the azimuthal crystallization pattern for the CoTe(1)-$Al_2O_3$ film. The azimuthal scan was taken as the film was rotated about the c-axis. Since the NiAs structure is six-fold symmetric, the appearance of the (103) peak every 30° indicates that there are two azimuthal stacking orientations in the film. The insert shows a $\theta/2\theta$-scan with the azimuthal angle set at the peak maximum in the azimuthal scan.

C. Energy-Dispersive X-Ray Spectroscopy

In order to obtain a quantitative estimate of the elemental Te/Co content, energy-dispersive x-ray spectroscopy (EDXS) was carried out using a JEOL JSM-6335F field emission scanning electron microscope (SEM). Figure 4(a) shows the EDXS spectrum for the CoTe(1)-MgO, $CoTe_{1.71}$, film. Figure 4(b) is an image of the film surface near its edge. Both the Co L-edge and the Te L-edge peaks are clearly visible in the spectrum indicating the presence of both elements in the film. A quantitative estimate of the atomic ratio Te/Co = 1.14 ± 0.12 for the CoTe(1)-MgO film was made by counting the fluorescence photons of L-edges for Co and Te at various points on the surface. However, this result might underestimate the amount of tellurium. Since the Co $L_\alpha$-edge (0.776 keV) and Te M-edge (0.778 keV) are so close, Te M-edge fluorescence might erroneously enhance the Co $L_\alpha$-edge spectrum [19].

D.  Magnetization

Measurements of the dc magnetization were carried out for magnetic fields −50 kOe ≤ H ≤ +50 kOe over the temperature range 5.0 K ≤ T ≤ 300 K using a superconducting quantum interference device (SQUID) MPMS-5 magnetometer from Quantum Design. The hysteresis loops were obtained with the magnetic field applied in the plane (parallel to the ab-plane) of the film; the films were mounted on a quartz rod for the parallel field measurements. In order to check for spurious background contributions, corresponding magnetic measurements were also made on clean substrates. As is the case with this work, great care must be taken to avoid contamination when making measurements involving nanoscale magnetism (e.g., moment values < $10^{-4}$ emu) [20]. Furthermore, the SQUID response curves are somewhat distorted as the samples are in the form of finite planes and not ideal dipoles [21]. Finally, magnetic artifacts can occur if the sample chamber contains even small amounts of residual oxygen [22]. These issues, all of which are important for the work reported here, are discussed in detail in [20-22].

Figure 5(a) shows the magnetic field dependence of the total magnetic moment −50 kOe ≤ H ≤ +50 kOe at the temperature T = 300 K for the CoTe(1)-MgO film (open circles) and the CoTe(2)-MgO film (closed circles), with the magnetic field applied in the ab-plane (perpendicular to the c-axis) of the films. It can be seen that, a small, but clear, ferromagnetic component appears for the CoTe(2)-MgO film with a much larger ferromagnetic component for the CoTe(1)-MgO film. In order to isolate the ferromagnetic component due to the film, linear fits were made to the high-field (|H| ≥ 10 kOe) portions of each curve shown in Fig. 5(a), which accounts for the diamagnetic susceptibility of the substrate plus a very small paramagnetic contribution due to the film. These corrections were then subtracted from the curves in Fig. 5(a) to obtain the curves shown in Fig. 5(b). For comparison, consistent correction values were also obtained by directly measuring the magnetic field dependence and temperature dependence of the magnetization for the various clean substrates. In order to make a quantitative comparison between the two films, the film ordered moment obtained after the subtraction was divided by the effective volume of the film. However, since the films are two phase, "arbitrary units" are used in Fig. 5(b). From Fig. 5(b), it can be seen that the ferromagnetic component for the CoTe(1)-MgO film is approximately four times that for the CoTe(2)-MgO film. As described below, NMR clearly identifies the ferromagnetic component as metallic Co. For the CoTe(1)-MgO film, the Co ferromagnetic moment value is 0.0005 emu as seen in Fig. 5(a). Using the saturation magnetization for bulk Co ($\approx$ 160 emu/g), an estimate of the mass density for $\beta(Co_2Te_3)$ ($\approx$ 8 g/cm$^3$), and the approximate volume of the film ($\approx 3.8 \times 10^{-6}$ cm$^3$), it is estimated that 10% of the mass of the film is due to the metallic Co. Furthermore, it follows that the Co:Te ratio is 47:53, which compares well with the PLD target ratio of 50:50.

E. Nuclear Magnetic Resonance

"Zero-field" $^{59}$Co spin-echo NMR spectra were obtained over the frequency range from $\nu$ = 210 MHz to 235 MHz using a Matec model 7700 pulsed-oscillator mainframe with a model 765 pulsed-oscillator/receiver, with the sample in a tuned circuit that was matched to 50 ohms. The NMR echo amplitude was optimized using a standard $\tau_{p1}$-$\tau$-$\tau_{p2}$ spin-echo pulse sequence with $\tau_{p1}$, $\tau_{p2}$ = 5.0 µs rf pulses, a pulse separation of $\tau$ = 20 µs, and a repetition rate of 33 Hz. Spectra were obtained by averaging the NMR signals 500 to 1,000 times with a 1.0 MHz interval across the frequency range. A resolution of 0.20 MHz is consistent with the 5.0 µs rf pulses. Spin-echo NMR spectra were obtained at T = 4.2 K with H = 0 and H = 8.0 kOe. The NMR sensitivity of the spectrometer (including the $1/\nu^2$ correction) was monitored over the entire frequency range by injecting an rf calibration pulse signal using a 50 ohm antenna. Measurements of the spin-spin relaxation time $T_2$ were made at selected frequencies across the spectrum by varying the pulse separation time from $\tau$ = 15 µs to 45 µs. As discussed below, the frequency dependence of $T_2$ can result in a significant correction to the NMR spectrum [23]. Operation at liquid He temperature was carried out using a conventional glass double dewar system.

As mentioned above, the motivation for the NMR experiments was to confirm the existence and identify the structure of the Co metal phase in the CoTe$_x$ films. The Co metal phase was indicated by the appearance of a ferromagnetic contribution in the SQUID magnetic measurements. A serious challenge existed in the NMR experiments due to the extremely small NMR coil filling factor. In addition, the Co metal occurs as a "trace amount" or "second phase" in the films. The observation of the $^{59}$Co signal is only possible due to the existence of the NMR enhancement factor which occurs in ferromagnetically- and ferrimagnetically-ordered materials [24]. In these experiments, the four-turn NMR coil had a rectangular cross-section 3.0 mm × 6.0 mm and was 6.0 mm in length. Five 5.0 mm × 5.0 mm CoTe(1) films on substrates were stacked tightly inside the coil.

Figure 6 shows the $^{59}$Co spin-echo NMR spectrum obtained from the stacked CoTe(1) films at 4.2 K for both H = 0 (closed circles) and 8.0 kOe (open circles) external magnetic field. The spectrum for H = 0 is quite broad, suggesting more than one component, and has a maximum at approximately 223 MHz. The single peak positions for face-centered-cubic (fcc) and hexagonal close-packed cobalt (dw = domain wall, d = domain, and np = nanoparticle) are indicated. The spectrum for H = 8.0 kOe is also broad with twin-peak-like features at 216 MHz and 221 MHz. The overall spectrum appears to be shifted downward in frequency by approximately 2.6 MHz. The application of the external magnetic field reduces the NMR echo amplitude considerably; however, the spectrum for H = 8.0 kOe has been normalized to that for H = 0 in Fig. 6 in order to compare the details. Figure 7 shows the NMR spin-echo amplitude at 220 MHz as a function of the applied magnetic field. (220 MHz is at a relatively flat part of the spectrum and hence, the reduction in echo amplitude is not due to the shift.) The magnetic field behavior is consistent with the reduction of the NMR enhancement factor which characterizes domains. These measurements clearly show that metallic cobalt appears to exist in the films.

Measurements of the spin-spin relaxation time $T_2$ were made at selected frequencies across both spectra. The variation of the echo amplitude with the rf pulse separation was always a single exponential. For the H = 0 spectrum, values of $T_2$ = 110 µs, 75 µs, 70 µs, and 64 µs were obtained for ν = 215 MHz, 220 MHz, 225 MHz, and 230 MHz, respectively, while for H = 8.0 kOe, $T_2$ = 99 µs, 130 µs, 74 µs for ν = 215 MHz, 220 MHz, and 225 MHz, respectively. Since $T_2$ is relatively long compared to the rf pulse separation for spectrum acquisition (20 µs), the correction is minimal.

## III. DISCUSSION AND CONCLUSIONS

Concerning previous work on the CoTe$_x$ system, two issues need to be addressed. The first issue is whether or not the stoichiometric CoTe composition actually occurs; i.e. does the system form the "perfect" NiAs structure with all of the Co sites filled in the middle planes between the Te atoms (see Fig. 1). As discussed above, it has been suggested that CoTe$_x$ for $0 \leq x \leq 1.20$ might be a eutectic mixture of metallic Co and the nonmagnetic compound CoTe$_{1.20}$. In support of this picture, Uchida [7] has made detailed magnetic measurements which are compelling. On the other hand, there are reports in the literature of materials with the CoTe stoichiometry, also having the NiAs structure with somewhat smaller a and c lattice parameters than those for CoTe$_{1.20}$ [3,15,18]. The second issue concerns the source of the magnetism observed in recent studies on nanostructured CoTe$_x$ materials [9-13]. Wang et al. [1] carried out high-resolution x-ray photoelectron spectroscopy (XPS) on cobalt telluride branched nanostructured samples. They observed weak satellite peaks in their Co $2p_{3/2}/2p_{1/2}$ spectra which they attribute to elemental Co [1]. Wu et al. [5] observed second phase fcc Co peaks in their XRD patterns from carbon supported CoTe$_{1.20}$ nanoparticles; the second phase peaks were reduced for higher heat treatment temperatures.

This report presents a structural and magnetic study of two sets of epitaxial CoTe$_x$ films (x = 1.71 and 1.76) prepared by PLD using two targets with different Co:Te atomic ratios. While XRD did not see the cobalt phase in the films, $^{59}$Co NMR clearly identifies metallic Co as the magnetic second phase (see Fig. 6). Although the existence of some amount of cobalt oxide phases(s) cannot be completely ruled out, the oxides tend to be antiferromagnetic with a completely different NMR signature. Furthermore, the calculation above with the magnetic data indicates that essentially all of the Co is accounted for in the films. Figure 7 shows a reduction in the NMR signal intensity with the application of a magnetic field. Based on a consideration of the (high) rf power level, the magnetic field behavior suggests that a significant portion of the NMR signal arises from domains, as would be the case for nanoparticles, and not domain walls, which characterize multidomain or bulk-like particles [24]. If this is the case, the Co inclusions in the films would be smaller than 76 nm, which is the approximate domain size in metallic Co [25]. This would also provide a possible explanation for the absence of metallic Co lines in the XRD scans. While both the fcc and hcp phases can exist at room temperature, the fcc structure is preferred above 450 °C with the hcp structure preferred below 450 °C [26]. However, the fcc structure has appeared for nanoscale particles at room temperature and below [26].

In order to identify crystallographic phase(s) from the $^{59}$Co NMR spectra, an assignment of the spectral features must include both structural and magnetic considerations. Recently, a comprehensive discussion of the various spectral features for both fcc and hcp Co has been provided by Andreev et al. [27]. For reference, the NMR peak frequencies (at T = 4.2 K and H = 0) for bulk (domain wall) fcc Co (217.0 MHz), nanoparticle (domain) fcc Co (220.8 MHz) and bulk (domain wall) hcp Co (228.0 MHz) are shown in Fig. 6 [28]. Although very weak and broad peaks sometimes occur due to twin and stacking faults, the fcc domain wall, fcc domain, and hcp domain wall cases for Co are each characterized by only a single main peak. Due to local field anisotropy, the situation for the hcp domain case is more complex [27]. However, based on experimental results for hcp nanoparticles, an assignment of 225.5 MHz is reasonable [29]. As can be seen in Fig. 6, the $^{59}$Co NMR spectrum for the CoTe(1) films is characterized by a broad peak with apparently two structural components; the components clearly do not match peaks associated with domain walls in the fcc or hcp phases. A more likely description is that the broad peak is characteristic of a combination of fcc and hcp domain peaks. The application of an external magnetic field results in a downward shift and narrowing of the component peaks; however, the separation remains consistent with the fcc and hcp domain frequencies. It is noteworthy that the epsilon cobalt (ε-Co) phase, which has been identified in nanoparticles as well as single crystals ≤ 0.3 μm, has two distinct Co sites [26]. There is one report of $^{59}$Co NMR in ε-Co; however, the spectra were obtained from single-domain 6.5 nm particles and, therefore, extremely broad, particularly on the low frequency side [23]. Although the ε-Co NMR spectra showed considerable intensity over the frequency range 220 MHz to 225 MHz, the association of the spectral features shown in Fig. 6 to ε-Co is not possible. In any case, $^{59}$Co NMR clearly identifies the existence of metallic cobalt as a second phase and a source of magnetism in the CoTe$_x$ films. On the other hand, XRD shows that the principal phase is hexagonal β(Co$_2$Te$_3$) which has the NiAs structure. This result suggests that recent reports of CoTe being a magnetic semiconductor should be revisited.

ACKNOWLEDGMENTS

The authors would like to thank Ch. Niedermayer for helpful discussions. Special thanks go to D. Morales for her help on the XRD measurements and R. Bibeault for his help on the NMR measurements. Work at the University of Connecticut was supported by DOE-BES Contract No. DEFG02-00ER45801.

FIGURES

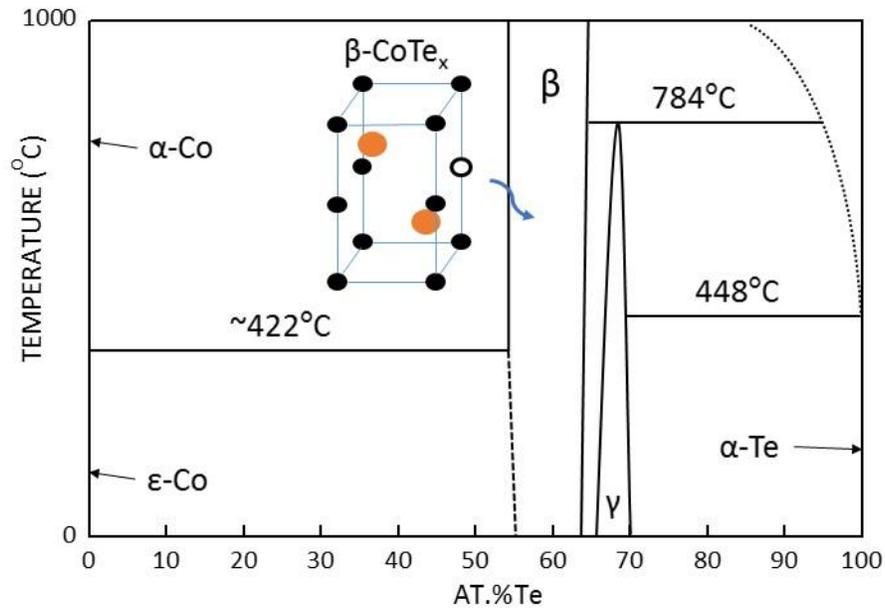

Fig. 1. (Color on-line) Phase diagram for the Co-Te system as assembled by K. Ishida and T. Nishizawa [14]. The hexagonal $\beta(Co_2Te_3)$ phase has the NiAs structure (space group $P6_3/mmc$) with a homogeneity range from 54.5 at.% Te ($CoTe_{1.20}$) to 64.3 at.% Te ($CoTe_{1.80}$). As the Te content increases from x = 1.20 to 1.80, Co atoms are removed continuously from the middle layer in the unit cell (see insert), eventually evolving into the orthogonal $\gamma(CoTe_2)$ phase.

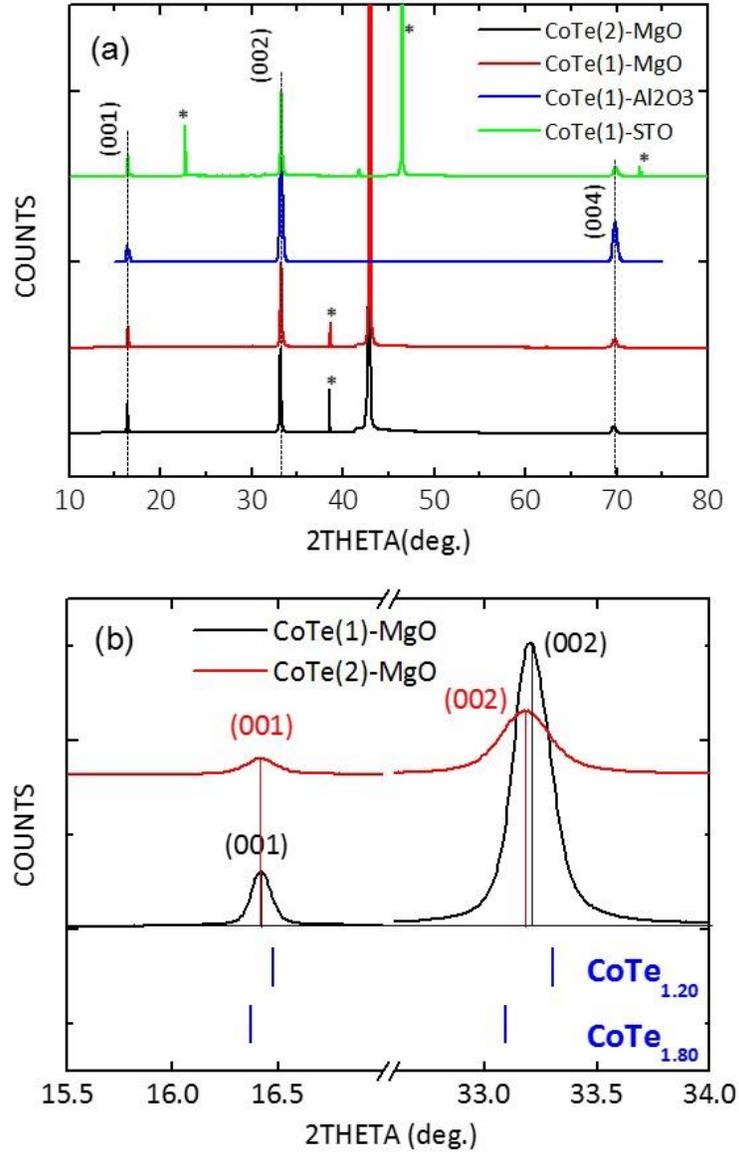

Fig. 2. (Color on-line) X-ray diffraction patterns for the CoTe$_x$ films. (a) The four representative epitaxial films are CoTe(1)-MgO, CoTe(1)-Al$_2$O$_3$, CoTe(1)-STO, and CoTe(2)-MgO, with the peaks being assigned to the (001), (002), and (004) reflections of the hexagonal NiAs structure which characterizes the β(Co$_2$Te$_3$) phase. "*" denotes the substrate peaks. (b) From the (002) peak positions for the CoTe(1)-MgO and CoTe(2)-MgO films, the compositions CoTe$_{1.71}$ and CoTe$_{1.76}$ are assigned, respectively, which are within the range for the β(Co$_2$Te$_3$) phase. The calculated peak positions for the β(Co$_2$Te$_3$) phase composition limits, CoTe$_{1.20}$ and CoTe$_{1.80}$, are indicated.

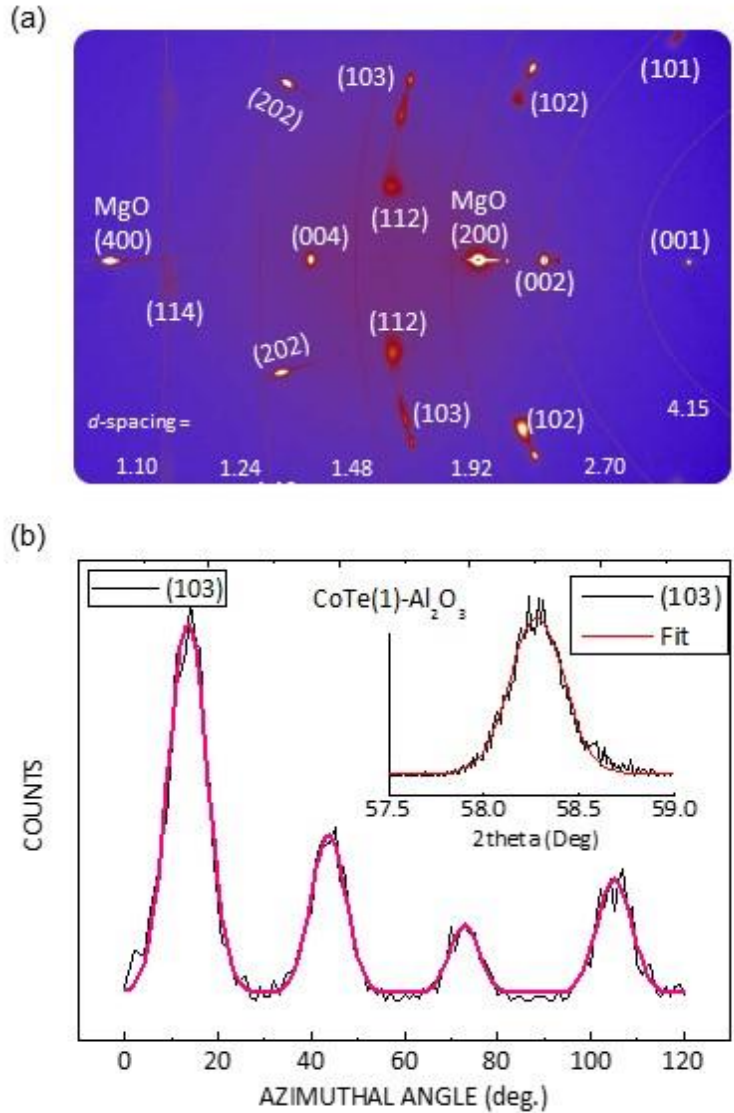

Fig. 3. (Color on-line) (a) X-ray diffraction patterns obtained from the 2D Oxford diffractometer for the CoTe(1)-Al$_2$O$_3$ film. The image was obtained from a θ-scan with the film aligned such that the c-axis stayed in the same plane as the incoming beam during sample rocking. From the image, all of the spots can be assigned to reciprocal lattice G-vector values characteristic of the hexagonal NiAs (P6$_3$/mmc) crystal structure [18]. (b) The azimuthal crystallization pattern for the CoTe(1)-Al$_2$O$_3$ film. The figure shows the azimuthal scan taken when the film was rotated about the c-axis. The inset shows a θ/2θ-scan with the azimuthal angle set at the peak maximum in the azimuthal scan.

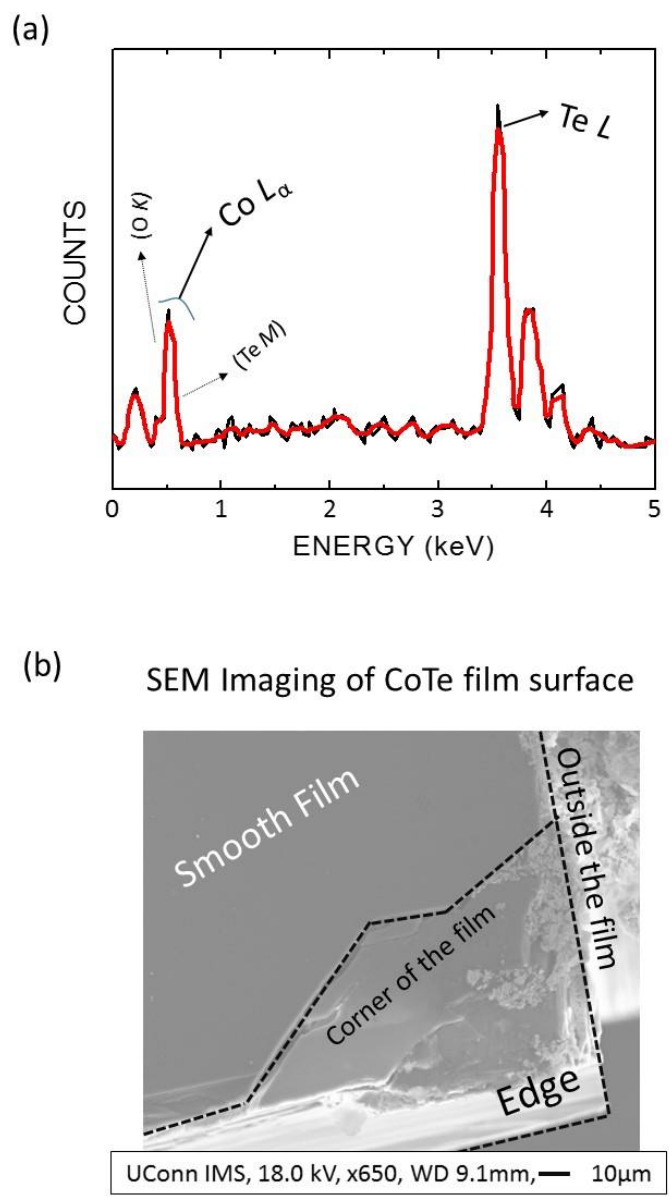

Fig. 4. (Color on-line) (a) Energy dispersive x-ray spectrum for the CoTe(1)-MgO, CoTe$_{1.71}$, film. The black curve is the experimental data and the red curve is a smoothed fit curve. Both the Co L-edge and the Te L-edge peaks are clearly visible; a quantitative estimate for the atomic ratio Te/Co = 1.14 ± 0.12 is obtained. (b) The image shows the surface of the film near its edge. Film surface, edge and outside the film are indicated with legends and the borders are shown with dashed lines. The corner is shown just for a visual comparison with the smooth film.

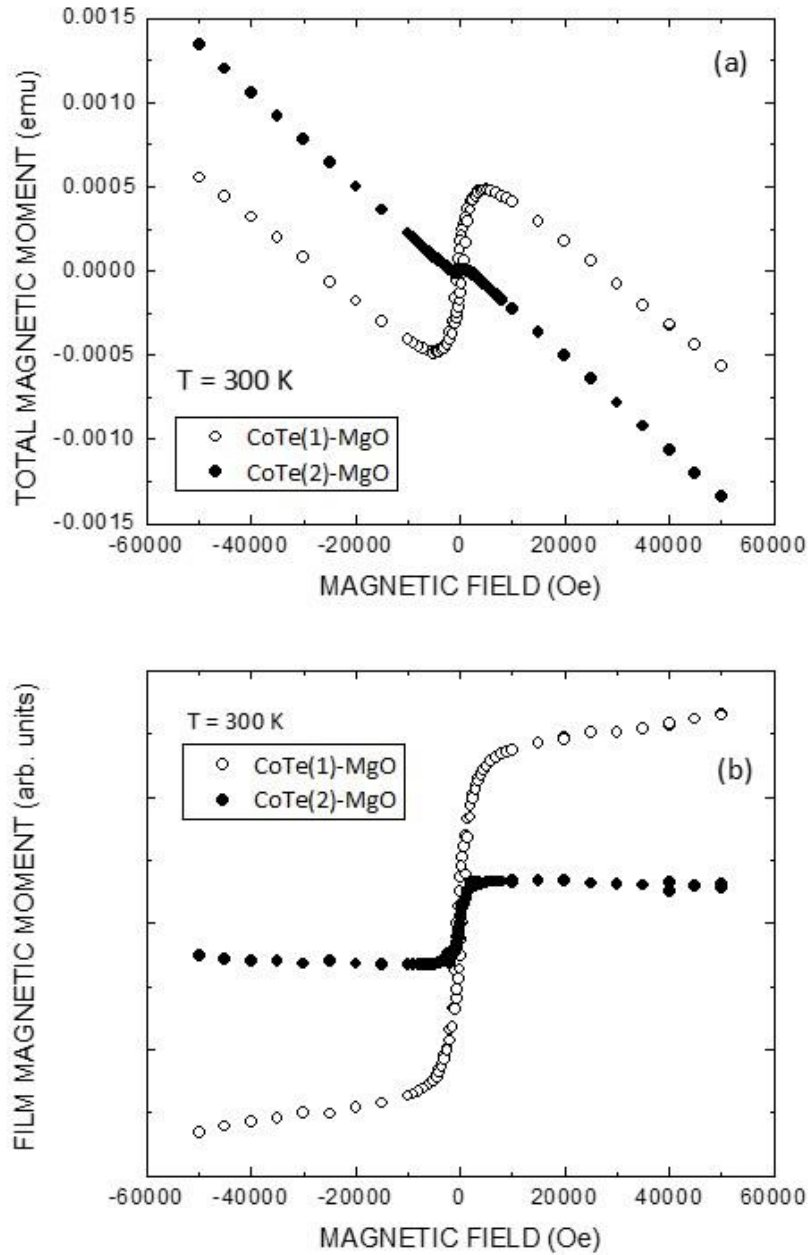

Fig. 5. (a) Total magnetic moment versus magnetic field at T = 300 K for the CoTe(1)-MgO film (open circles) and the CoTe(2)-MgO film (closed circles), with the magnetic field applied in the ab-plane (perpendicular to the c-axis of the films). (b) The magnetic field dependence (or hysteresis loop) of just the film (ferromagnetically-ordered) magnetization at T = 300 K which was obtained by subtracting the diamagnetic component from data in Fig. 5(a) and normalizing for the film volume. The ferromagnetic component for the CoTe(1)-MgO film is approximately four times that for the CoTe(2)-MgO film.

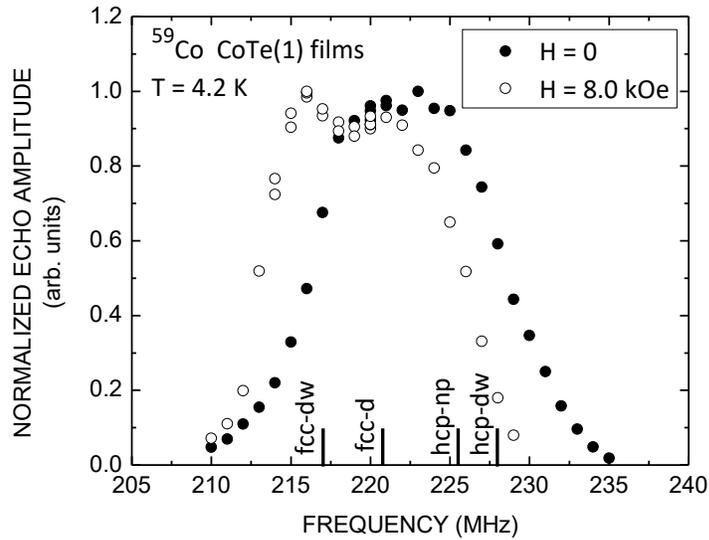

Fig. 6. $^{59}$Co spin-echo NMR spectrum obtained from the stacked CoTe(1) films at 4.2 K for both H = 0 (closed circles) and 8.0 kOe (open circles) external magnetic field. The spectrum for H = 0 is quite broad, suggesting more than one component, with a maximum at approximately 223 MHz. The single peak positions for fcc and hcp cobalt (dw = domain wall, d = domain, and np = nanoparticle) are indicated. The spectrum for H = 8.0 kOe is also broad with twin-peak-like features at 216 MHz and 221 MHz. The overall spectrum appears to be shifted downward in frequency by approximately 2.6 MHz. The application of the external magnetic field reduces the NMR echo amplitude considerably; however, the spectrum for H = 8.0 kOe has been normalized to that for H = 0 in order to compare the details.

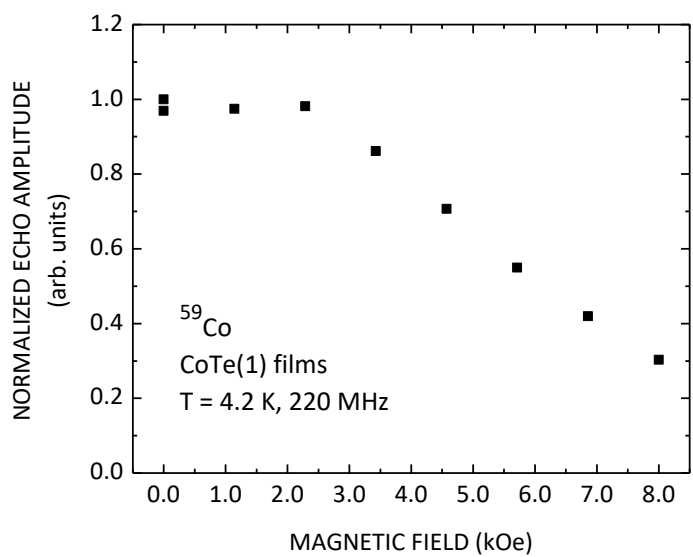

Fig. 7. The NMR spin-echo amplitude at 220 MHz as a function of the applied magnetic field. (220 MHz is at a relatively flat part of the spectrum and hence, the reduction in echo amplitude is not due to the shift.) The magnetic field behavior is consistent with a reduction of the NMR enhancement factor most likely within domains.